\documentclass[reprint,amsmath,amssymb,aps,]{revtex4-2}
\usepackage{verbatim}
\usepackage{graphicx}
\usepackage{dcolumn}
\usepackage{bm}
\usepackage{amsmath}
\usepackage{braket} 
\usepackage{ulem}
\usepackage[linkcolor=blue,colorlinks=true,breaklinks=true,citecolor=blue,urlcolor=blue]{hyperref}
\usepackage{booktabs} 
\usepackage{makecell} 
\usepackage{xcolor}

\begin{document}

\preprint{APS/123-QED}

\title{Valley-polarized Quantum Anomalous Hall and Topological Metal Phase in Rashba-induced pseudospin-1 lattice}
\author{Puspita Parui}
    \email{puspitaparui44@gmail.com}
\author{Bheema Lingam Chittari}%
    \email{bheemalingam@iiserkol.ac.in}
\affiliation{Department of Physical Sciences, Indian Institute of Science Education and Research Kolkata, Mohanpur 741246, West Bengal, India}

\begin{abstract}
We study the topological properties of Rashba spin-orbit coupling and exchange coupling induced pseudospin-$1$ system Dice lattice under the influence of a staggered electric potential and magnetization. The band structure and topological phases of the system are investigated and compared with the pseudospin-$\frac{1}{2}$ system honeycomb lattice. Under individual influence of the staggered electric field and magnetization, the system undergoes a distinct phase transition: (i) a staggered electric potential drives the system from a quantum anomalous Hall $(C_n = 2)$ to a valley polarized quantum anomalous Hall phase $(C_n = -1)$ associated with edge modes with a flip in the chirality; while (ii) a staggered magnetization changes the system to a topological metal associated with unconventional antichiral edge bands, from a topological insulator. These results are further supported by calculations of the Chern phase diagrams, Hall conductance, zigzag, and armchair edge states. Our findings enhance the understanding of new topological phases in the 2D pseudospin-1 system and open up a new platform to explore the anti-chiral edge states.
\end{abstract}

\maketitle

\section{\label{sec:level1}Introduction}
Two-dimensional (2D) materials with band structures possessing a non-trivial topology have gained enormous research interest since they were first introduced by Haldane~\cite{PhysRevLett.61.2015} that showed the quantized Hall conductance in the absence of an external magnetic field, which is known as the Quantum Anomalous Hall effect (QAH) and was later experimentally realized~\cite{PhysRevLett.101.246810, chang2013experimental,jotzu2014experimental}. Essentially, it is achieved by breaking the time-reversal symmetry (TRS) using a complex next-nearest-neighbor hopping with opposite phases for two different sublattices in the Honeycomb structure~\cite{PhysRevLett.61.2015}. In 2005, Kane and Mele ingeniously introduced a model that preserves time-reversal symmetry induced by strong intrinsic spin-orbit coupling (ISOC)~\cite{PhysRevLett.95.226801, PhysRevLett.95.146802}. It shows the quantum spin Hall (QSH) effect, and host helical edge states are protected by time-reversal symmetry and characterized by a $Z2$ topological invariant. As the QAH and QSH phase display dissipationless edge states and topological currents that are extremely robust against disorder effects~\cite{RevModPhys.82.3045}, they constitute a very attractive platform for ultralow-power electronics and spintronics~\cite{RevModPhys.82.3045, Ren_2016, qiao2011electronic}.
Recently, the Kane-Mele model of the Honeycomb lattice was generalized to $\alpha$-$\mathcal{T}_3$ lattice~\cite{PhysRevB.103.075419}, where the tunable hopping parameter $\alpha$ $(0\leq \alpha \leq 1)$ interpolates between the honeycomb $(\alpha=0)$ and Dice $(\alpha=1)$ lattice~\cite{PhysRevLett.81.5888, PhysRevLett.112.026402, PhysRevB.92.245410, biswas2016magnetotransport, PhysRevB.107.085408, PhysRevB.109.235105} and was found to undergo a topological phase transition at $\alpha=1/2$ from QSHI with $\sigma_{xy} = e^2/h$ $(\alpha=0)$ to $\sigma_{xy}=2e^2/h$ $(\alpha=1)$~\cite{PhysRevB.103.075419}.
The $\alpha$-$\mathcal{T}_3$ lattice is characterized by a flat band along with two dispersive bands forming a Dirac cone similar to the Honeycomb lattice, hence realized as a pseudospin-1 Dirac-Weyl system~\cite{PhysRevA.80.063603}. 
Due to the spin-valley splitting of energy bands for $0 < \alpha < 1$, the system depicts a significantly rich phase diagram when a staggered magnetization term breaks the TRS of the QSH phase~\cite{PhysRevB.103.075419}.
The property of the dispersionless zero-energy flat band and the variable Berry phase~\cite{PhysRevB.92.245410} leads to many unconventional phenomena, such as unconventional Hall effect~\cite{PhysRevB.92.245410, PhysRevB.96.155301, biswas2016magnetotransport}, higher Chern insulating phases~\cite{PhysRevB.107.035421, PhysRevB.101.235406, PhysRevB.109.165118, PhysRevB.110.045426}, unconventional Anderson-localization~\cite{PhysRevB.100.104201}, super-Klein-tunneling~\cite{PhysRevB.96.024304, PhysRevB.95.235432}, flat-band ferromagnetism~\cite{taie2015coherent}, unusual Landau-Zener Bloch oscillations,  and peculiar magnetic-optical effect~\cite{PhysRevB.94.125435}.
The $\alpha$-$\mathcal{T}_3$ lattice can naturally be built by growing a heterostructure of cubic Transition metal oxides ($\rm SrTiO_3/SrIrO_3/SrTiO_3$) or by creating an optical lattice with three pairs of counter-propagating laser beams~\cite{PhysRevLett.112.026402}. It has been recently shown that $\rm Hg_{1-x}Cd_xTe$ with critical doping can be mapped onto $\alpha$-$\mathcal{T}_3$ lattice with $\alpha=1/\sqrt3$~\cite{PhysRevB.92.035118}.
Many other aspects of $\alpha$-$\mathcal{T}_3$ lattice have been addressed so far, such as thermoelectric properties~\cite{Alam_2019}, optically irritated Floquet engineering~\cite{PhysRevB.99.205429, PhysRevB.111.045406}, Strain-induced pseudo magnetic field~\cite{PhysRevB.106.155417}, and the effect of Rashba spin-orbit coupling~\cite{PhysRevB.84.241103, lin2023interaction}. 
Although it is shown that extrinsic Rashba spin-orbit coupling (RSOC) is detrimental to the quantum spin Hall effect~\cite{PhysRevLett.95.146802}, it helps to realize the QAH effect. The QAH effect was predicted to be realized by introducing Rashba SOC and an exchange field in graphene~\cite{PhysRevB.82.161414, PhysRevB.83.155447, zhang2015quantum} and was experimentally observed in magnetic insulators~\cite {chang2013experimental}.  
In the Dice lattice, introducing an RSOC in the presence of a magnetic field~\cite{PhysRevB.84.241103} or Coulomb interaction~\cite{PhysRevB.102.045105}, it is possible to get nearly flat bands with $C=\pm2$. However, the fate of this quantum Hall phase in the presence of staggered potential or magnetization remains unexplored. In this paper, we aimed to study the possible topological phases in $\alpha$-$\mathcal{T}_3$ lattice, mainly focusing on the Dice lattice, which is the $\alpha=1$ limit of the general $\alpha$-$\mathcal{T}_3$ lattice. The emergent topological phases arise due to the interplay of RSOC and ferromagnetic exchange coupling (EC) with staggered electric potential and magnetization terms individually.
We carefully investigated the valley-polarized QAH (VPQAH) and argue that such staggered potentials are not always detrimental to the quantum Hall phases, comparing $\alpha$-$\mathcal{T}_3$ lattice (pseudospin-1) with that of the Honeycomb lattice (pseudospin-1/2) with similar effects. Further, we present the anti-chiral edge states (ACES) associated with the topological metal (TM) phase of $\alpha$-$\mathcal{T}_3$ lattice. Such ACES have recently been studied extensively in single-layer~\cite{PhysRevLett.120.086603, PhysRevB.109.235105} as well as composite forms of the modified Haldane model~\cite{PhysRevB.110.165303} and are experimentally evident~\cite{PhysRevLett.125.263603}.

\begin{figure}[!ht]
    \centering
    \includegraphics[scale=1.2]{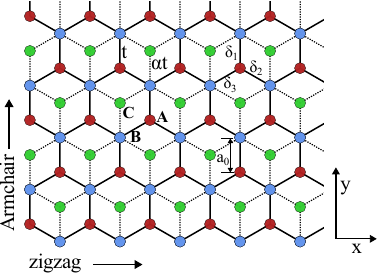}
    \caption{Schematic of the $\alpha$-$\mathcal{T}_3$ lattice with zigzag and armchair edges. $\delta_n (n=1,2,3)$ are three nearest neighbors drawn from the rim sublattice ($A$ or $C$). $A$,$B$, and $C$ are denoted by the colors red, blue, and green, respectively.}
    \label{fig:lattice_schematic}
\end{figure}

The rest of the paper is arranged as follows. In Sec.\ref{sec:model_discussion}, we introduced the tight-binding model Hamiltonian for the $\alpha$-$\mathcal{T}_3$ lattice with RSOC, exchange coupling, and external potentials. In Sec.\ref{sec:results_and_discussions}, we discussed the topological phase transitions with external potential and magnetization, which includes bulk band structure, topological invariant, Hall conductance, and edge modes of ribbon geometry, and conclude in Sec.\ref{sec:conclusions}

\section{System Hamiltonian and Formalism} \label{sec:model_discussion}
The $\alpha$-$\mathcal{T}_3$ lattice with Hamiltonian with Rashba spin-orbit coupling and exchange term can be written as 
\begin{equation}
    H = H_0 + H_R + H_{ex}
    \label{eq:Full_hamiltonian}
\end{equation}
where the first term
\begin{equation}
    H_0 = - t\sum_{<i,j>,s} C_{i,s}^{\dagger} C_{j,s}- \;\alpha t \sum_{<j,k>,s} C_{j,s}^{\dagger} C_{k,s} + H.c.
\end{equation}
describes the nearest neighbor (NN) hopping along directions $\boldsymbol{\delta_1} = (0,a_0)$, $\boldsymbol{\delta_2} = (-\frac{\sqrt{3}a_0}{2}, -\frac{a_0}{2})$, and $\boldsymbol{\delta_3} = (\frac{\sqrt{3}a_0}{2}, -\frac{a_0}{2})$, from $B$ and $A(C)$ sites, with hopping strength $t (\alpha t)$ with spin index $s$. Here, $C_{i,s}^\dagger (C_{i,s})$ are electron creation (annihilation) operator with spin polarization $s$ acting on site $i$. $a_0$ is the distance between two sublattices. The second term 

\begin{multline}
H_R = i\lambda_R \left[ \sum_{<i,j>,s,s^\prime} \hat{\bm{e}}_z \cdot (\bm{\sigma}_{s,s^\prime} \times \bm{d}_{i,j}) {C_{i,s}}^\dagger C_{j,s^\prime} \right. \\
\left. + \;\alpha \sum_{<j,k>,s,s^\prime} \hat{\bm{e}}_z \cdot (\bm{\sigma}_{s,s^\prime} \times \bm{d}_{j,k}) {C_{j,s}}^\dagger C_{k,s^\prime} \right] +H.c.
\end{multline}

describes the spin-mixing Rashba spin-orbit coupling with coupling strength $\lambda_R$ $(\alpha \lambda_R)$ \cite{PhysRevB.110.235432, islam2024screw}, while $\bm{\sigma}$ $(\bm{\sigma}_x, \bm{\sigma}_y, \bm{\sigma}_z)$ are Pauli matrices and $\bm{d}_{ij}$ represents a unit vector pointing from site $j(k)$ to site $i(j)$ and finally the ferromagnetic exchange term modeled by 
\begin{equation}
    H_{ex} = \sum_{i,s} \bm{\sigma}_z \lambda_{ex} C^\dagger_{i,s}  C_{i,s} + H.c.
\end{equation}
Throughout this article, we measure the Rashba SOC and exchange energy in units of the hopping parameter $t$.

The momentum space Hamiltonian in the sublattice basis $\{ \ket{A_{\uparrow}},\ket{B_{\uparrow}},\ket{C_{\uparrow}},\ket{A_{\downarrow}},\ket{B_{\downarrow}},\ket{C_{\downarrow}} \}^T$ obtatined by fourier transforming Eq.\eqref{eq:Full_hamiltonian} can be described as, 
\begin{equation}
    H(\boldsymbol{k}) =  \begin{pmatrix}
        H_\uparrow & H_{\uparrow\downarrow}\\
        {H_{\downarrow\uparrow}} & H_\downarrow
    \end{pmatrix}
    \label{eq:H_k_full}
\end{equation}

where, $H_\uparrow(H_\downarrow)$ is the spin-up (spin-down) Hamiltonian and $H_{\uparrow\downarrow}(H_{\downarrow\uparrow})$ is the spin-mixing part of the  Hamiltonian.
\begin{equation}
    H_\uparrow(H_\downarrow) =  \begin{pmatrix}
        +(-)\lambda_{ex} & f(\bm{k},t) & 0\\
        f^*(\bm{k},t) & +(-)\lambda_{ex} & f(\bm{k},\alpha t)\\
        0 & f^*(\bm{k},\alpha t) & +(-)\lambda_{ex}
    \end{pmatrix}    
\end{equation}

\begin{equation}
    H_{\uparrow\downarrow} =  \begin{pmatrix}
        0 & \rho(\bm{k}) & 0\\
        -\rho(-\bm{k}) & 0 & -\alpha\rho(\bm{k})\\
        0 & \alpha\rho(-\bm{k}) & 0
    \end{pmatrix}    
\end{equation}
and, $H_{\downarrow\uparrow} = H_{\uparrow\downarrow}^\dagger $
where, $f(\bm{k},t^\prime) = f_x(\bm{k},t^\prime) - if_y(\bm{k},t^\prime)$, and $t^\prime$ takes the values of either $t$ or $\alpha t$.

\begin{multline}
f_x(\bm{k},t^\prime) = t^\prime \bigg\{\cos \left(a_{0} k_y \right) \\
+ 2 \cos \left( \frac{\sqrt{3} a_{0} k_x}{2} \right)
\cos \left( \frac{a_{0} k_y}{2} \right)
\bigg\},
\end{multline}

\begin{figure*}[!ht]
    \centering
    \includegraphics[scale=0.3]{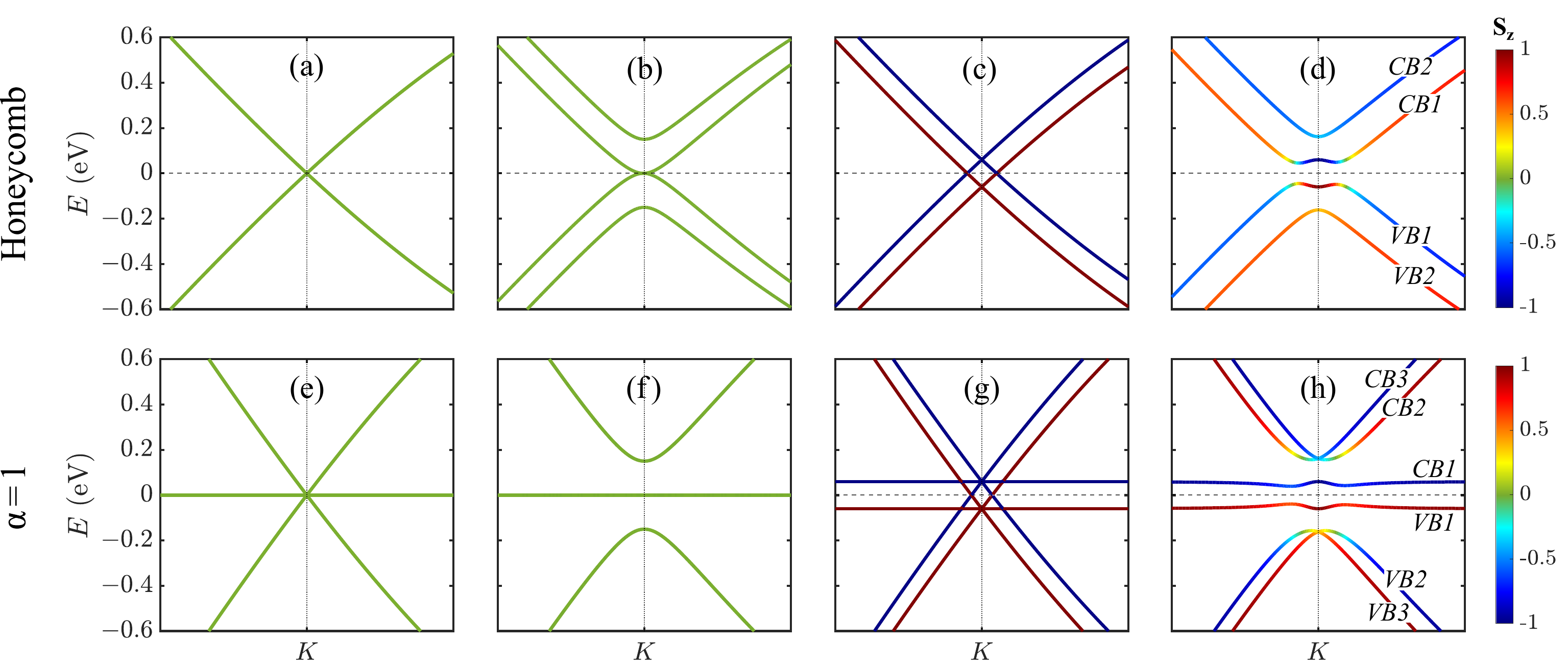}
    \caption{Band structure of the system around valley $K$. The upper and lower panel corresponds to $\alpha=0$ (honeycomb lattice) and $\alpha=1$ (dice lattice), respectively. The column represents the results for $\lambda_R = 0$, $\lambda_{ex} = 0$ [panels (a) and (e)], $\lambda_R \neq 0$, $\lambda_{ex} = 0$ [panels (b) and (f)], $\lambda_R = 0$, $\lambda_{ex} \neq 0$ [panels (c) and (g)] and $\lambda_R \neq 0$, $\lambda_{ex} \neq 0$ [panels (d) and (h)].The band color depicts the $z$-component of the spin texture. The tight binding parameters used in the unit of $t$ are $\lambda_R = 0.05$, $\lambda_{ex} = 0.06$.}
    \label{fig:band_structure_spin_polarized}
\end{figure*}

\begin{multline}
    f_y(\bm{k},t^\prime) = t^\prime\bigg\{\sin \left(a_{0}k_y\right)\\
    - 2\cos \left( \frac{\sqrt{3} a_{0}k_x}{2} \right)
     \sin \left( \frac{a_{0}k_y}{2} \right)
     \bigg\},
\end{multline}

\begin{multline}
    \rho(\bm{k}) = i\lambda_R \bigg\{ e^{-ia_0k_y}\; + \\
     \; 2 \; e^{\frac{ia_0k_y}{2}}\cos{\left( \frac{\sqrt{3}a_0k_x}{2} + \frac{2\pi}{3} \right)} \bigg\}
\end{multline}

\begin{multline}
    \rho(-\bm{k}) = i\lambda_R \bigg\{ e^{ia_0k_y} \; + \\
     2 \; e^{-\frac{ia_0k_y}{2}}\cos{\left( \frac{\sqrt{3}a_0k_x}{2} - \frac{2\pi}{3} \right)} \bigg\}
\end{multline}
The topological invariant associated with the system is calculated using Berry curvature for individual bands in the $k_x-k_y$ plane~\cite{RevModPhys.82.1959}. The z-component of the berry curvature was calculated numerically using~\cite{RevModPhys.82.1959}
\begin{equation}
    \Omega_n(k_x,k_y) = -2\sum_{n^\prime \ne n} \text{Im} \bigg[\frac {\bra{u_n} \frac{\partial H}{\partial k_x} \ket{u_{n^\prime}} \bra{u_{n^\prime}} \frac{\partial H}{\partial k_y} \ket{u_n}} {(E_{n^\prime} - E_n)^2} \bigg]
    \label{eq:bc_equation}
\end{equation}
for $n$th band for each $k$-points, summing over all other bands $n^{\prime}$. $\ket{u_n}$ and $E_n$ are block states and eigenvalues of the Hamiltonian given in Eq.(\ref{eq:Full_hamiltonian}) for the $n$th band. The surface integral of the Berry curvature $\Omega_n$ over the first Brillouin zone gives $2\pi C_n$, where $C_n$ is called the Chern number or TKNN index~\cite{PhysRevLett.49.405, PhysRevLett.51.51} of $n$th band, which is the topological invariant of the system. 

\section{Results and Discussion} \label{sec:results_and_discussions}

\subsection{ Momentum interlocked spin polarization of bands}
The bulk band structure of $\alpha$-$\mathcal{T}_3$ lattice with Rashba SOC and exchange coupling can be obtained by diagonalizing $6 \times 6$ spin-full Hamiltonian of Eq.\eqref{eq:H_k_full} for each crystal momentum. 
In Fig.~\ref{fig:band_structure_spin_polarized}, we presented the bulk energy spectrum at $K$ - valley for two limiting cases, $\alpha=0$, which is Honeycomb lattice (pseudospin-$\frac{1}{2}$) in the upper panel and $\alpha=1$, which is the Dice lattice (pseudospin-$1$) in the lower panel. The expectation value of the $z$-component of the spin is shown as the band color. Our results for $\alpha=0$ are consistent with earlier studies~\cite{PhysRevB.82.161414}.
In the absence of spin-orbit coupling and exchange term,  ($\lambda_R = 0$, $\lambda_{ex} = 0$), due to spin degeneracy, the bands are four-fold for $\alpha = 0$ and six-fold for $\alpha = 1$ degenerate respectively near the Dirac points
as shown in panels (a) and (e) of Fig.~\ref{fig:band_structure_spin_polarized}.
The inclusion of Rashba SOC term,($\lambda_R  \neq 0$, $\lambda_{ex} = 0$) couples the electron's spin to its momentum in the plane and breaks the inversion symmetry in the case of in $\alpha=0$, leading to the splitting of the bands into four non-degenerate bands by an amount proportional to Rashba coupling strength $\lambda_R$ \cite{PhysRevLett.95.226801, PhysRevB.82.113405, PhysRevB.80.235431}. In the case of $\alpha=1$, though a spin-momentum coupling arises due to Rashba SOC, however, the inversion symmetry is still preserved~\cite{PhysRevB.108.075166}, and this allows the three spin-degenerate bands to remain degenerate with their corresponding spin-flipped counterparts. Panels (b) and (f) show the corresponding band structures. The exchange field alone ($\lambda_R = 0$, $\lambda_{ex} \neq 0$) makes the bands fully spin-polarized as shown in panels (c) and (g). The spin-up (spin-down) bands are pushed upward (downward) by an amount of exchange field, and the time-reversal symmetry (TRS) is broken due to the spin polarization. The simultaneous presence of RSOC and exchange field ($\lambda_R \neq 0$, $\lambda_{ex} \neq 0$) makes the bands nondegenerate, and a bulk gap opens up near the Dirac points. 
Due to the interplay of RSOC and EC, a momentum-dependent spin texture is observed near the band-crossing points. The corresponding band structures are shown in panels (b) and (h). 
The bulk gaps obtained in both Honeycomb and Dice lattice Fig.~\ref{fig:band_structure_spin_polarized} are found to be topological in nature with a total Chern number of occupied bands, $|C|=2$ and exhibit a QAH phase with $\sigma_{xy} = 2e^2/h$ ~\cite{PhysRevB.82.161414, PhysRevB.84.241103}. The Chern numbers of individual bands are $C_\text{VB1(CB1)}=\mp 1$ and $C_\text{VB2(CB2)}=\pm3$ for the Honeycomb and $C_\text{VB1(CB1)}=\pm 2$, $C_\text{VB2(CB2)}=\pm1$ and $C_\text{VB3(CB3)}=\mp1$ for the Dice lattice. Although the $\alpha$ plays an important role in tuning the spin Hall phase in the case of ISOC~\cite{PhysRevB.103.075419}, it does not change the QAH phase obtained by RSOC and EC. The impact of $\alpha$ on the spectral properties, along with different RSOC and EC strengths, is shown in Appendix~\ref{app:intermediate_alpha}. However, the bulk gaps are sensitive to external staggered fields that can even modify their existing QAH phases, as we will see in the later sections. 

\subsection{Staggered electric field induced valley polarized quantum anomalous Hall}
\label{sec: b}
In this section, we investigate the effect of a staggered electric potential on the  
Rahsba SOC and exchange field induced dice lattice and compare the results with that of the honeycomb lattice. Here, for the Dice lattice, we consider a staggered electric potential $\Delta$, i.e., the electric potential on A and C sublattices are opposite but zero in the B sites. Therefore, we add another term $H_\Delta = \sum_{i,s} \Delta_i  C^\dagger_{i,s}  C_{i,s} $ in the Hamiltonian of Eq.(\ref{eq:Full_hamiltonian}), which in sublattice basis reads as,
\begin{equation*}
    H_\Delta = \Delta\sigma_0 \otimes
\begin{pmatrix}
   1 & 0 & 0\\
   0 & 0 & 0\\
   0 & 0 & -1   
\end{pmatrix} = \Delta\sigma_0 \otimes S_z
\end{equation*}
 where $\sigma_0$ is the identity matrix in the spin basis and $S_z$ is the z-component of pseudospin-$1$ matrix. This staggered potential term $S_z$ is well known for gap generating with an effective mass at the Dirac points \cite{PhysRevB.96.024304, PhysRevB.103.195442} and for modulating the band gap for a spin-Hall system \cite{PhysRevB.103.075419}. Another possible staggered potential term, such as a certain $U$-term used in the Hamiltonian of Ref.\cite{PhysRevB.96.024304, PhysRevB.103.195442} shifts the quasi-flat middle bands (VB1 and CB1) upward (downward) for $+U(-U)$ in our system and does not lead to the band gap modulation, which is our main focus. Therefore, we restrict ourselves to the inclusion of the $S_z$-term in our Hamiltonian. Physically, the onsite energy can be tuned in a system of ultracold atoms forming an optical Dice lattice\cite{PhysRevB.73.144511, PhysRevA.80.063603} by controlling the intensity and detuning the laser beams \cite{tarruell2012creating} forming the lattice. Another possible physical platform is the $\text{SrTiO3/SrIrO3/SrTiO3}$ trilayer heterostructure grown along the $(111)$ direction \cite{PhysRevB.84.241103}. Since the three sublattices (A, B, C) lie on three distinct planes, a staggered A-C potential can be generated using a dual gate (top and bottom) arrangement such that the hub-site B (middle plane) lies on the zero-potential reference point.
For the Honeycomb lattice, such a staggered potential is conventionally realized by taking $+\Delta$ on the $A$ sublattice and $-\Delta$ on the $B$ sublattice. The staggered electric potential (${\Delta}$) breaks the inversion symmetry, and the bands are no longer symmetric under the valley exchange. Topological phase transitions involve bulk gap closing and reopening, which in our system occurs at the Dirac points. Thus, analyzing band gaps solely at these points suffices to determine phase transitions. The  energies of the bands at ${K} (-\frac{2\pi}{3\sqrt{3} a_0}, \frac{2\pi}{3a_0})$ and ${K}^\prime (\frac{2\pi}{3\sqrt{3} a_0}, \frac{2\pi}{3a_0})$ for Honeycomb lattice are $E_\eta^\xi = -\xi (\Delta + \eta\lambda_{ex})$ and $E_\eta^{\prime\xi} = \xi \sqrt{(\Delta - \eta\lambda_{ex})^2 + 9\lambda_R^2}$ and for Dice lattice are $E_\eta^\xi = -\xi (\Delta + \eta\lambda_{ex})$, $E_\eta^{\prime\xi} = \frac{\xi}{2}\big(- \Delta + \sqrt{(\Delta - 2\eta\lambda_{ex})^2 + 36\lambda_R^2}\;\big )$ and  $E_\eta^{\prime\prime\xi} = \frac{\xi}{2}\big(+\Delta + \sqrt{(\Delta - 2\eta\lambda_{ex})^2 + 36\lambda_R^2}\; \big )$ with $\eta=+ (-)$ representing the $K (K^\prime)$ and $\xi=+(-)$ for CB (VB) respectively.

The bulk gap at the Dirac points calculated as $\big|\Delta E_\eta^{h(d)}\big| = \big|\textbf{min}[E_\eta^{+}, E_\eta^{\prime +}, E_\eta^{\prime\prime +}] - \textbf{max}[E_\eta^{-}, E_\eta^{\prime -}, E_\eta^{\prime\prime -}]\big|$, depends on the relative strengths of $\lambda_R$, $\lambda_{ex}$, and $\Delta$, where the superscript $h(d)$ denotes honeycomb (dice) lattice. Table-\ref{tab:table1} summarizes the relevant band gaps considering the minimum and maximum energy eigenvalues at the Dirac points.
As we vary $\Delta$, for both honeycomb and dice lattice, the bulk gap $\Delta E_{K^\prime}^{h(d)}$ first closes at $\Delta = \lambda_{ex}$ and then reopens for $\Delta > \lambda_{ex}$ at ${K}^\prime$, suggesting a topological phase transition (TPT) irrespective of Rashba strength. For Dice lattice, an additional TPT associated with $\Delta E_K^d=0$ is possible at $K$, contingent on the interplay between $\lambda_R$, $\lambda_{ex}$, and $\Delta$. However, this phase turns out to be trivial as we have shown later in the phase diagram of Fig.\ref{fig:rsoc_delta_phase}.  In contrast, the honeycomb lattice exhibits no such phase transition at the $K$-point for any finite positive values of $\Delta,$ $\lambda_R$ and $\lambda_{ex}$.

\begin{table}[b]
\centering
\caption{\label{tab:table1}
Bulk gap at the Dirac points for the Honeycomb and the Dice lattice depending on the relative strength of $\lambda_R$, $
\lambda_{ex}$ and $\Delta$. Case-I: $9\lambda_R^2 < 4\Delta\lambda_{ex}$ $ (< 2\Delta^2 + 4\Delta\lambda_{ex})$ and case-II: $9\lambda_R^2 > 4\Delta\lambda_{ex}$ $( > 2\Delta^2 + 4\Delta\lambda_{ex})$ for Honeycomb (dice) lattice}

\medskip
\begin{tabular*}{\columnwidth}{@{\extracolsep{\fill}}lcc@{}}
\toprule
\textbf{Gaps} & \textbf{case-I} & \textbf{case-II}\\

\midrule
$\Delta E_{K^\prime}^{h(d)}$ & $E_{K^\prime}^+ - E_{K^\prime}^-$ & $E_{K^\prime}^+ - E_{K^\prime}^-$\\
$\Delta E_{K}^{h(d)}$ & $E_{K}^+ - E_{K}^-$ & $E_{K}^{\prime +} - E_{K}^{\prime -}\;$ \\

\bottomrule
\end{tabular*}
\end{table}

\begin{figure}[!ht]
    \centering
    \includegraphics[scale=0.33]{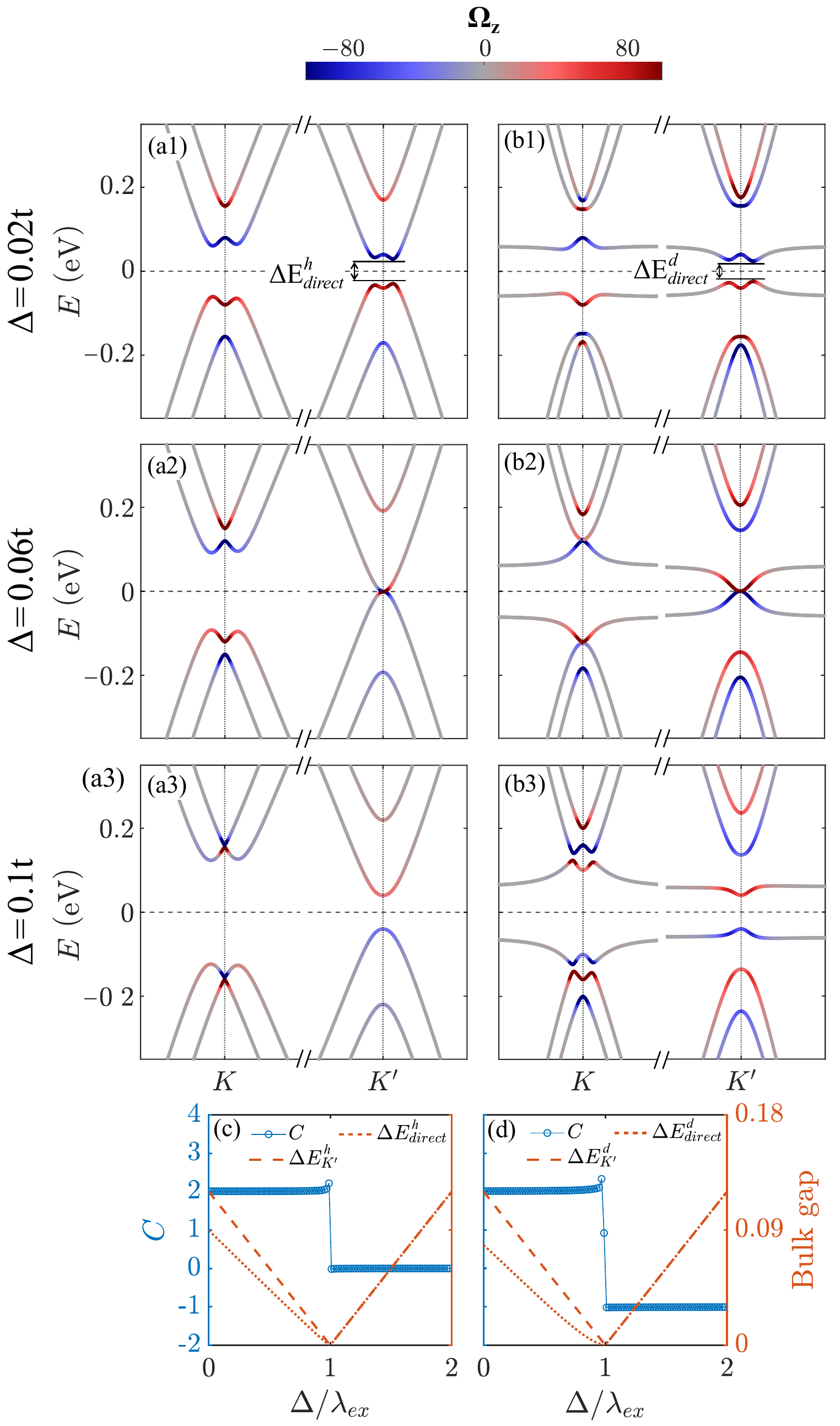}
    \caption{Band structures of the gapped phase in the presence of RSOC and exchange coupling are shown for different values of $\Delta$ near the two Dirac points $K$ and $K^\prime$ for $(\alpha=0)$ honeycomb lattice [(a1)-(a3) of left panel] and $(\alpha=1)$ Dice lattice [(b1)-(b3) of right panel]. Band color represents the $z$ component of the berry curvature. The total Chern number of all occupied bands and the bulk band gap as a function of staggered potential for the Honeycomb lattice (c) and Dice lattice (d). The parameters used are the same as that of Fig.\ref{fig:band_structure_spin_polarized}}
    \label{fig:bulk_bs_bgap_chocc_delta}
\end{figure}

\begin{figure}[!ht]
    \centering
    \includegraphics[scale=0.4]{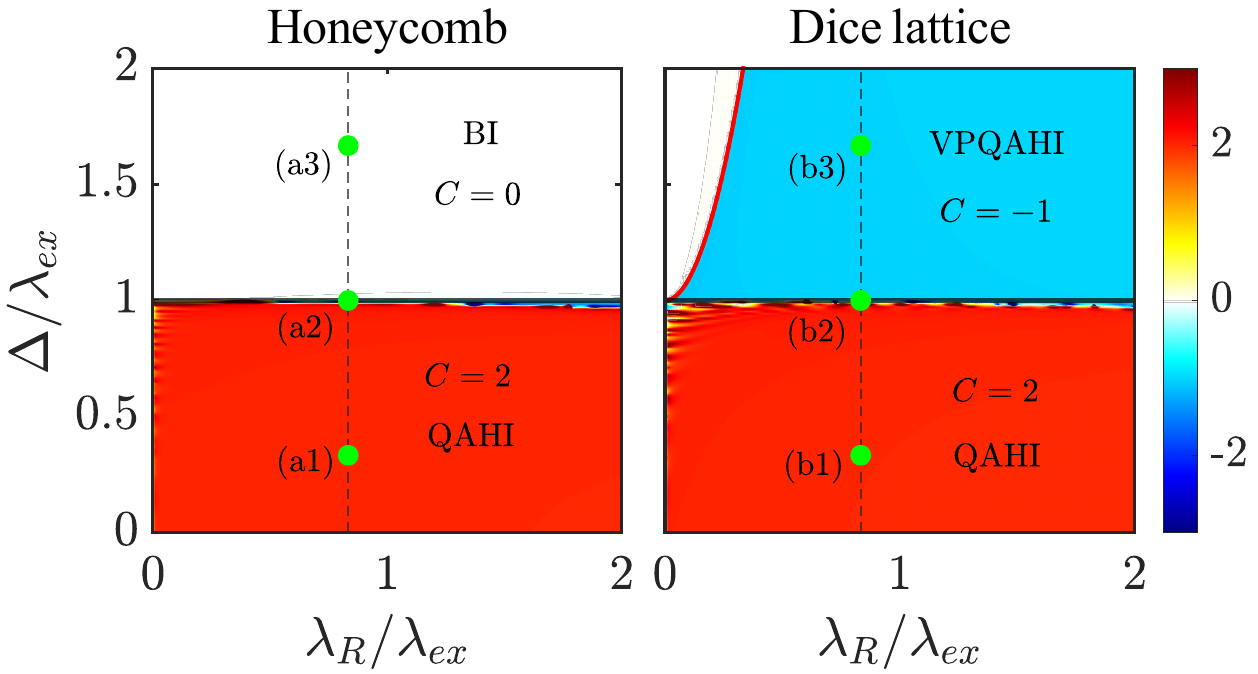}
    \caption{Chern phase diagram in the parameter space $\big(\lambda_R/\lambda_{ex}, \Delta/\lambda_{ex}\big)$, where $C$ represents the total Chern number of all occupied bands. The red and black phase boundaries are obtained via $\Delta E_{K}^{h(d)}=0$ and $\Delta E_{K^\prime}^{h(d)}=0$, respectively. Green dots indicate parameter sets corresponding to the band structures in Fig.\ref{fig:bulk_bs_bgap_chocc_delta}. The black dashed line corresponds to the $\lambda_R/\lambda_{ex}$, kept fixed throughout the manuscript. The other parameters are the same as in Fig.\ref{fig:band_structure_spin_polarized}.
    }
    \label{fig:rsoc_delta_phase}
\end{figure}

In Fig.\ref{fig:bulk_bs_bgap_chocc_delta} (a1)-(a3) and (b1)-(b3) we show the bulk band structure near the two inequivalent valleys for the Honeycomb [left panel] and Dice lattice [right panel], respectively, for three different values of staggered potential before the transition $(\Delta = 0.02t < \lambda_{ex})$, at the critical point $(\Delta = 0.06t = \lambda_{ex})$ and after the transition $(\Delta = 0.1t > \lambda_{ex})$.

The band color represents the $z$-component of berry curvature, where the change in polarity of berry curvature at the band crossings suggests a topological phase transition for individual bands. The plot of bulk bandgap $(\Delta E_{direct}^{h(d)})$, {bulk gap at $K^\prime$ ($\Delta E_{K^\prime}^{h(d)}$)} and total chern number $(C)$ of occupied bands as a function of potential strength $\Delta$ are shown in Figs.\ref{fig:bulk_bs_bgap_chocc_delta}(c) and \ref{fig:bulk_bs_bgap_chocc_delta}(d). The bulk gap $\Delta E_{direct}^{h(d)}$ is numerically calculated as $\text{\textbf{min}(CB1)} - \text{\textbf{max}(VB1)}$ and is different from $\Delta E_{K^\prime}^{h(d)}$, but they merge at the band crossing point $K^\prime$.
Although Fig.\ref{fig:bulk_bs_bgap_chocc_delta} shows multiple band crossings among the higher energy bands away from the Fermi energy, the total Chern number below the Fermi energy remains the same and only the bulk gap closing and opening modifies the topological phase of the system very similar to a hidden topological phase transition shown recently in Honeycomb lattice~\cite{PhysRevB.110.165303}.
The band crossing at the bulk gap modifies the topology of the two systems differently. For the Honeycomb lattice, the topological phase goes from a higher Chern insulator (HCI) with $C = 2$ to a bulk insulator (BI) phase with $C = 0$ as shown by the blue curve in Fig.\ref{fig:bulk_bs_bgap_chocc_delta}(c). Whereas for the dice lattice, the phase goes from HCI ($C = 2$) to another Chern insulating phase with $C = -1$ as shown by the blue curve in Fig.\ref{fig:bulk_bs_bgap_chocc_delta}(d). The negative sign implies a switch in the chirality of the edge modes associated with the Hall phase, which we have verified by numerically calculating the edge bands later in Sec.\ref{sec:edge_bands}. The edge modes associated with the $C = -1$ phase arise from one specific valley instead of equal contribution from each valley as in the case of the conventional QAH phase ($C=2$ region). Such unequal valley contribution of edge modes leads to valley-polarized QAH phase (VPQAH) \cite{PhysRevLett.112.106802, PhysRevB.101.155425,PhysRevB.104.L161113} and we have discussed this elaborately in Sec.\ref{sec:edge_bands}. Therefore, by applying an external staggered electric potential of strength equal to the exchange coupling, one can switch the chirality of topologically protected edge states in a pseudospin-1 system, like a dice lattice. Changing the sign of the electric potential can reverse the valley polarization and chirality of the edge states. 

We further investigate the sensitivity of these topological phases to the Rashba SOC strength $\lambda_R$. Fig.\ref{fig:rsoc_delta_phase} presents the Chern phase diagram, where the color of the surface plot indicates numerically computed total Chern number as a function of $\Delta/\lambda_{ex}$ and $\lambda_R/\lambda_{ex}$. The honeycomb lattice exhibits no dependence on $\lambda_R$, and its phase diagram solely reflects the topological phases described in Fig.\ref{fig:bulk_bs_bgap_chocc_delta}. However, the dice lattice reveals an additional trivial phase region with $C=0$ at low Rashba for $\Delta/\lambda_{ex} >1$. The phase boundaries in the diagram correspond to bulk gap closing at the $K$ and $K^\prime$ points, obtained via $\Delta E_{K (K^\prime)}^{h(d)} = 0$. The black solid line $(\Delta=\lambda_{ex})$ in both diagrams indicates the phase boundaries corresponding to the phases discussed above and shown in Fig.\ref{fig:bulk_bs_bgap_chocc_delta}, obtained by $\Delta E_{K^\prime}^{h(d)}=0$. 
For Dice lattice, the bulk band gap at $K$ vanishes at $(\Delta/\lambda_{ex}) = 1+9(\lambda_R/\lambda_{ex})^2$ indicated as red solid curve in the phase diagram and obtained via $\Delta E_K^d =0$. This phase boundary separates the trivial region $(C=0)$ from the VPQAH region $(C=-1)$. The dashed line corresponds to the Rashba strength we kept fixed throughout the manuscript.

\begin{figure}[!ht]
    \centering
    \includegraphics[scale=0.33]{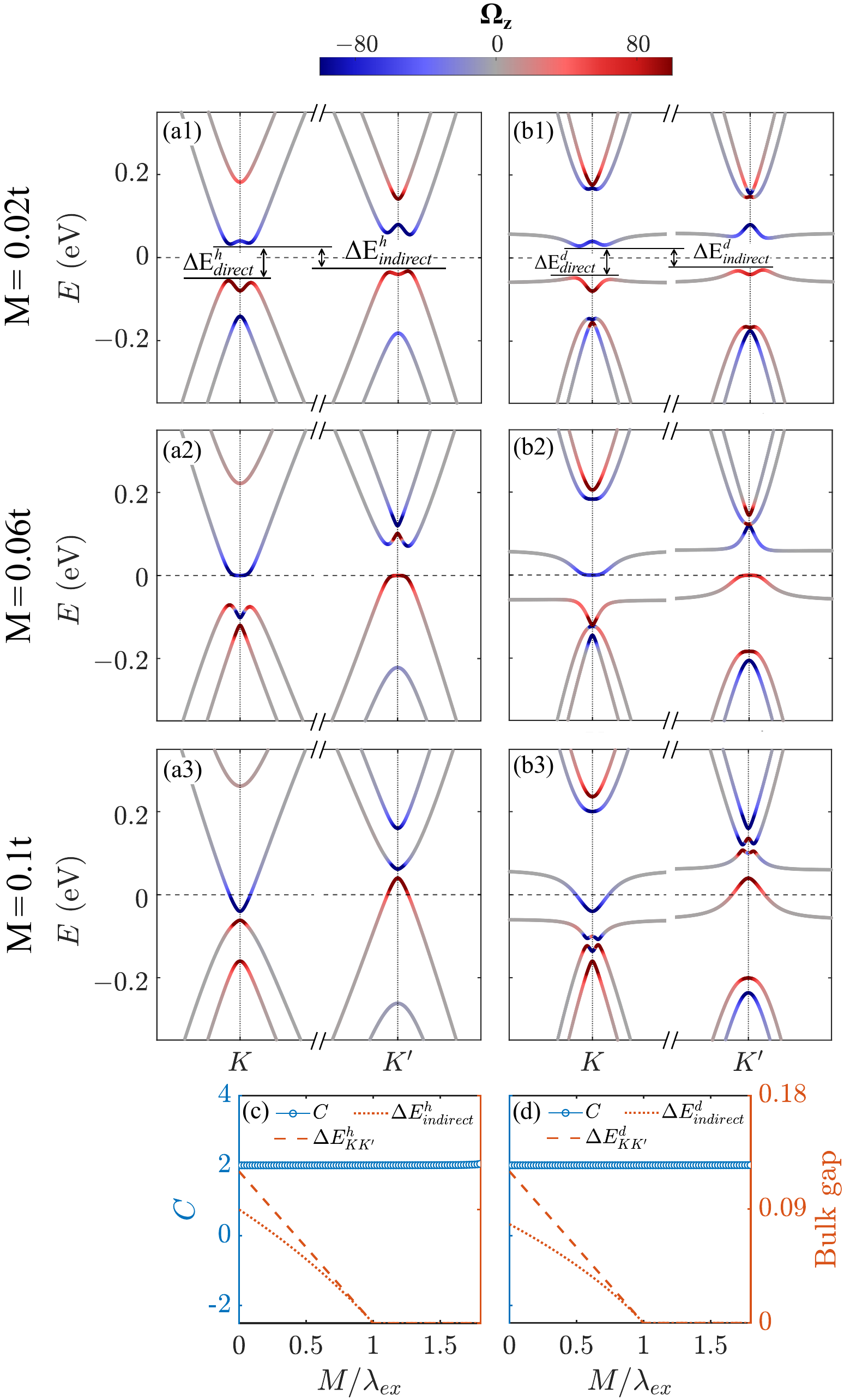}
    \caption{Band structures of the gapped phase in the presence of RSOC and exchange term are shown near the two Dirac points $K$ and $K^\prime$ for different values of staggered magnetization $M$ for honeycomb lattice  [(a1)-(a3) of left panel] and $(\alpha=1)$ Dice lattice  [(b1)-(b3) of right panel]. Band color represents the $z$ component of the berry curvature. The total Chern number ($C$) of all occupied bands and the bulk band gap as a function of staggered potential for the Honeycomb lattice (c) and Dice lattice (d). The parameters used are the same as that of Fig.\ref{fig:band_structure_spin_polarized}}
    \label{fig:bulk_bs_bgap_chocc_stmag}
\end{figure}

\begin{figure}[!ht]
    \centering
    \includegraphics[scale=0.4]{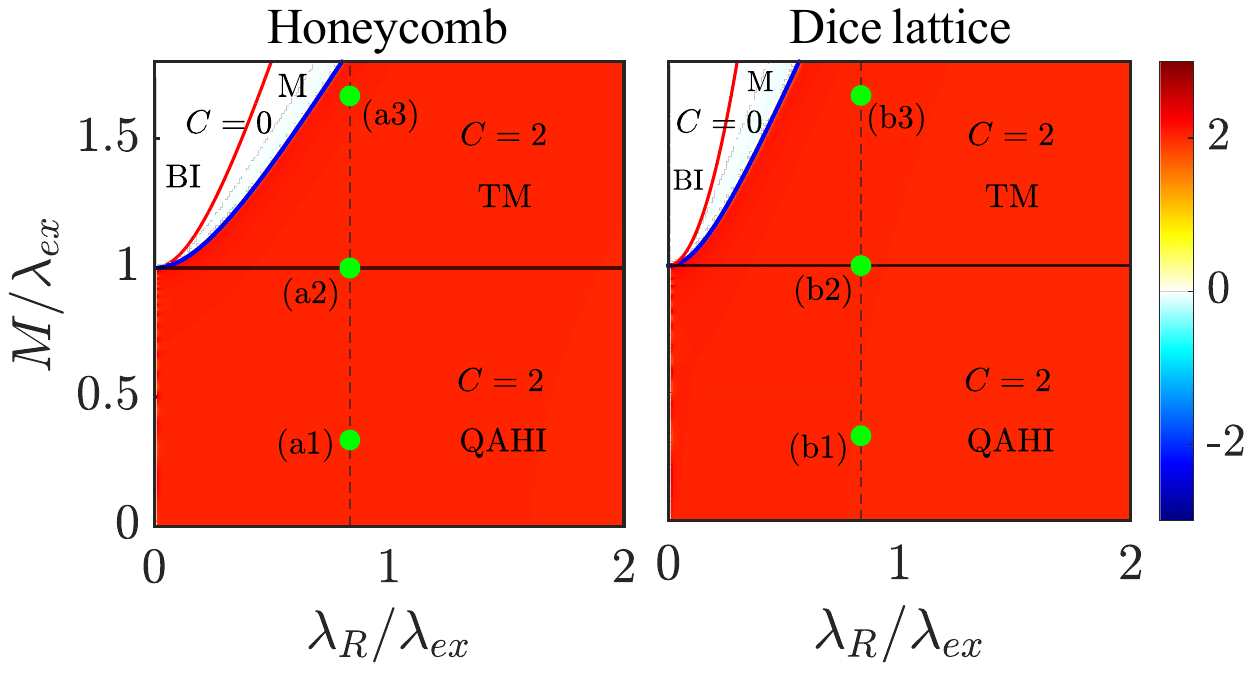}
    \caption{Chern phase diagram in the parameter space $\big(\lambda_R/\lambda_{ex}, M/\lambda_{ex}\big)$ where, $C$ represents the total Chern number of all occupied bands. The black and Blue phase boundaries are obtained via $\Delta E_{KK^\prime}^{h(d)}=0$ and $\Delta E_{K(K^\prime)}^{h(d)}=0$, respectively. The red phase boundary obtained via $\textbf{min}[E_K^{+},E_K^{\prime+},E_K^{\prime\prime+}] > 0~ (\textbf{max}[E_{K^\prime}^{-},E_{K^\prime}^{\prime-},E_{K^\prime}^{\prime\prime-}] > 0 )$) at $K \;(K^\prime)$. Green dots indicate parameter sets corresponding to the band structures in Fig.\ref{fig:bulk_bs_bgap_chocc_stmag}. The black dashed line corresponds to the Rashba strength we kept fixed throughout the manuscript. The other parameters are the same as in Fig.\ref{fig:band_structure_spin_polarized}.
    }
    \label{fig:rsoc_M_phase}
\end{figure}
\subsection{Staggered magnetization induced topological metal (TM) phase}
\label{sec: c}
Here, we discuss the effect of a staggered magnetization along with the RSOC and EC. We consider an $A-C$ sublattice staggered magnetization~\cite{PhysRevB.103.075419} $M$ by adding a term $H_M = \sum_{i,s} \sigma_z M_i C_{i,s}^\dagger C_{i,s}$ to the original Hamiltonian Eq.(\ref{eq:Full_hamiltonian}), which in sublattice basis reads as,
\begin{equation*}
    H_M = M \sigma_z \otimes
\begin{pmatrix}
   1 & 0 & 0\\
   0 & 0 & 0\\
   0 & 0 & -1   
\end{pmatrix} = M\sigma_z \otimes S_z
\end{equation*}

where $\sigma_z$ is the $z$-component of the Pauli matrix in the spin basis. As discussed in Sec.\ref{sec: b} an alternative $U$-term can also be used instead of $S_z$ matrix in the view of gap-generation or gap modulation \cite{PhysRevB.96.024304, PhysRevB.103.195442}. As a spin-dependent magnetization term, this indeed modulates the bulk gap; however, this drives our QAH system into a trivial insulator, which lacks significant physical interest for the present study. We therefore focus exclusively on the $S_z$ matrix as a staggered magnetization term. Experimentally, this implementation can be realized via the magnetic proximity effect in a $(111)$ oriented $\text{SrTiO3/SrIrO3/SrTiO3}$ trilayer heterostructure\cite{PhysRevB.84.241103}, sandwiching between a Ferromagnetic and an antiferromagnetic substrate.
For the Honeycomb lattice, such a staggered magnetization is implemented with an additional $M\sigma_z \tau_z$ term to the Hamiltonian~\cite{PhysRevB.87.155415}.
Unlike staggered potential, this term breaks the spectral symmetry with respect to the $E=0$ plane. Instead, the energy spectrum obeys $E(k) = -E(-k)$, which enforces an indirect bulk band gap.
The band energies at $K$ and $K^\prime$ for Honeycomb lattice are $E_\eta^\xi = (\eta M - \xi\lambda_{ex})$ and $E_\eta^{\prime\xi} = -\eta M +\xi \sqrt{\lambda_{ex}^2 + 9\lambda_r^2}$ and for Dice lattice are $E_\eta^\xi = \eta (M+\xi\lambda_{ex})$, $E_\eta^{\prime\xi} = \frac{1}{2}\big(-\eta M +\xi \sqrt{(M - 2\lambda_{ex})^2 + 36\lambda_R^2}\;\big)$ and $E_\eta^{\prime\prime\xi} = \frac{1}{2}(-\eta M +\xi \sqrt{\big(M + 2\lambda_{ex})^2 + 36\lambda_R^2}\;\big)$ with $\eta=+ (-)$ representing the $K (K^\prime)$ and $\xi=+(-)$ for CB (VB) respectively.
The relevant band gaps are are summarized in Table-\ref{tab:table2}, where, $\big|\Delta E_{KK^\prime}^{h(d)}\big|$ is the indirect bandgap measured from the Dirac points and calculated as, $\big|\Delta E_{KK^\prime}^{h(d)}\big| = \big|\textbf{min}[E_K^{+}, E_K^{\prime +}, E_K^{\prime\prime +}] - \textbf{max}[E_{K^\prime}^{-}, E_{K^\prime}^{\prime -}, E_{K^\prime}^{\prime\prime -}]\big|$ and $\big|\Delta E_{K(K^\prime)}^{h(d)}\big|$ is the direct bandgap at $K(K^\prime)$, calculated as described in the previous section. Due to the spectral symmetry $\big|\Delta E_K^{h(d)}\big| = \big|\Delta E_{K^\prime}^{h(d)}\big|$. As the parameter 
$M$ is varied in both the honeycomb and dice lattices, the indirect bandgap $\Delta E_{KK^\prime}^{h(d)}$ closes at $M=\lambda_{ex}$, regardless of the Rashba strength, signaling a transition to a metallic phase. This phase transition is not associated with a band crossing as in the case of the staggered potential described in the previous section (Fig.\ref{fig:bulk_bs_bgap_chocc_delta}). Additionally, Table~\ref{tab:table2} reveals another TPT associated with the vanishing of the direct bandgap at both $K$ and $K^\prime$ points ($\Delta E_{K(K^\prime)}^{h(d)}=0$) for both lattices, which is sensitive to the Rashba coupling.
\begin{table}[b]
\centering
\caption{\label{tab:table2} 
Bulk gap at the Dirac points for the Honeycomb and Dice lattice depending on the relative strength of $\lambda_R$, $\lambda_{\text{ex}}$, and $\Delta$. Case-I: $9\lambda_R^2 < 4M^2 + 4M\lambda_{\text{ex}}$ $ \big(< 2M^2 + 2M\lambda_{ex}\big)$ and case-II: $9\lambda_R^2 > 4M^2 + 4M\lambda_{\text{ex}}$ $ \big(> 2M^2 + 2M\lambda_{ex}\big)$ for Honeycomb (dice) lattice}

\medskip
\begin{tabular*}{\columnwidth}{@{\extracolsep{\fill}}lcc@{}}
\toprule
\textbf{Gaps} & \textbf{case-I} & \textbf{case-II} \\
\midrule
$\big|\Delta E_{KK'}^{h(d)}\big|$ & $E_K^+ - E_{K^\prime}^{-}$ & $E_K^+ - E_{K^\prime}^{-}$ \\[0.5em]

$\big|\Delta E_{K(K')}^{h(d)}\big|$ & $E_K^+ - E_{K}^{-}$ & $E_K^{+}-E_K^{\prime-}$ \\
\bottomrule
\end{tabular*}
\end{table}

Fig.\ref{fig:bulk_bs_bgap_chocc_stmag}(a1) - \ref{fig:bulk_bs_bgap_chocc_stmag}(a3) and \ref{fig:bulk_bs_bgap_chocc_stmag}(b1) - \ref{fig:bulk_bs_bgap_chocc_stmag}(b3) shows the bulk band structure near the two inequivalent Dirac points of Honeycomb lattice [left panel) and dice lattice [right panel] under the influence of three different magnetization strengths $M = 0.02t (< \lambda_{ex})$, $M = 0.06t (= \lambda_{ex})$, and $M = 0.1t (> \lambda_{ex})$ respectively. The color of the bands indicates the berry curvature polarization and distribution along the $k$-path.
The relevant bulk band gaps ($\Delta E_{indirect}^{h(d)}$), $\Delta E_{KK^\prime}^{h(d)}$ and the total Chern number $(C)$ are plotted with magnetization strength $M$ in Fig.\ref{fig:bulk_bs_bgap_chocc_stmag}(c) and \ref{fig:bulk_bs_bgap_chocc_stmag}(d). 
As the magnetization strength increases, the conduction band at the $K$ point and the valence band at the $K^\prime$ point shift toward the Fermi energy and touch the Fermi energy at $M = \lambda_{ex}$ (\ref{fig:bulk_bs_bgap_chocc_stmag}(a2) and \ref{fig:bulk_bs_bgap_chocc_stmag}(b2)) leading to the the bulk indirect band gap $\Delta E_{indirect}^{h(d)}$ (red dotted curve) and the $K$-$K^\prime$ gap $\Delta E_{KK^\prime}^{h(d)}$ (red dashed curve) to be vanished, rendering the system metallic. At $M = \lambda_{ex}$, these two gaps become equal.
As the bands are associated with non-zero Chern numbers and no band crossing between the valence and conduction bands is observed, we anticipate this new metallic phase to be a topological metal (TM), conducting phase hosting the co-existence of in-gap bulk states with edge states. Later in Sec.\ref{sec:edge_bands}, we discussed the nanoribbon band structures and confirmed that the system experiences a phase transition from HCI ($C=2$) to a TM at $M=\lambda_{ex}$, A similar phase transition has previously been observed in $\alpha$-$\mathcal{T}_3$ lattice at $\alpha=1/2$ for a Modified Haldane model~\cite{PhysRevB.109.235105}. We found that the staggered magnetization term is detrimental to the QAH phase for both Honeycomb and Dice lattices; however, it opens up a new platform to study unconventional edge band structures.

These topological metal phases exhibit a strong dependence on Rashba SOC strength. To characterize this behavior, we plotted a Chern phase diagram in the parameter space $\lambda_R/\lambda_{ex}$ and $M/\lambda_{ex}$, as shown in Fig.\ref{fig:rsoc_M_phase}. The color scale represents the total Chern number, summing contributions from VB1 and VB2 for the honeycomb lattice and from VB1, VB2, and VB3 for the dice lattice. The solid black line marks the quantum anomalous Hall insulator to topological metal (QAHI-TM) phase transition boundary, determined by the condition $\big|\Delta E_{KK^\prime}^{h(d)}\big|=0$ and consistent with the results shown in Fig.~\ref{fig:bulk_bs_bgap_chocc_stmag}. Notably, since the indirect band gap closing occurs without band crossing, the total Chern number remains unchanged ($C=2$) across both QAHI and TM phases. The white region in the phase diagram corresponds to a total Chern number $C=0$, which encompasses two distinct phases, A metal (M) phase characterized by $C=0$ with $\Delta E_{KK^\prime}^{h(d)}=0$ and a bulk insulator with $C=0$ with $\Delta E_{KK^\prime}^{h(d)}\neq0$. This region is separated from $C=2$ region by a blue solid phase boundary determined by the conditions, $(M/\lambda_{ex}) = \frac{1}{2}\big(1+\sqrt{1+9(\lambda_R/\lambda_{ex})^2}\big)$ and $(M/\lambda_{ex}) = \frac{1}{2}\big(1+\sqrt{1+18(\lambda_R/\lambda_{ex})^2}\big)$ for honeycomb and dice lattice respectively. These boundaries are derived from the direct band gap closing condition $\big|\Delta E_{K(K^\prime)}^{h(d)}\big|=0$. 
The red solid curve demarcating the boundary between the metal and BI within $C=0$ region is determined by $(M/\lambda_{ex}) = \sqrt{1+9(\lambda_R/\lambda_{ex})^2}$ for the honeycomb lattice and $(M/\lambda_{ex}) = 1+9(\lambda_R/\lambda_{ex})^2$  for the dice lattice. These phase boundaries can be derived from either $\textbf{min}(E_K^{+},E_K^{\prime+},E_K^{\prime\prime+}) > 0$ or $\textbf{max}(E_{K^\prime}^{-},E_{K^\prime}^{\prime-},E_{K^\prime}^{\prime\prime-}) > 0$, which physically signify reopening of the bulk gap, indicating a transition from metallic to an insulatting phase. The dashed line represents the fixed Rashba strength adopted throughout our manuscript.

\subsection{Chiral and Half antichiral edge states} \label{sec:edge_bands}
\begin{figure*}[!t]
    \centering
    \includegraphics[scale=0.4]{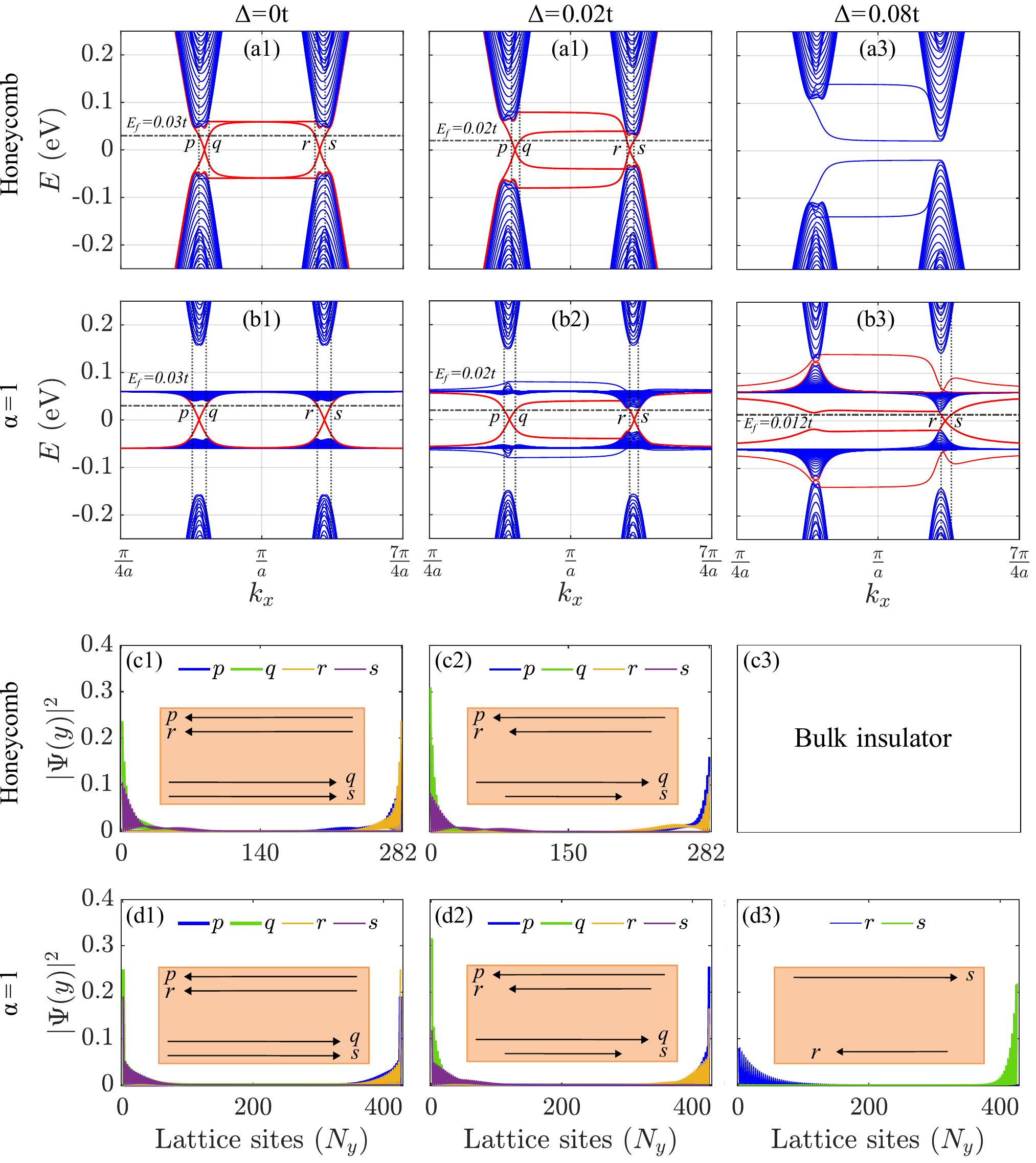}
    \caption{(upper panel) The band structures of zigzag nanoribbon with applied Rashba spin-orbit coupling and exchange term for Honeycomb lattice $[(a1) - (a3)]$ and dice lattice $[(b1) - (b3)]$ for three different values of staggered electric potential $\Delta$, where red and blue color indicates the conducting edge bands and bulk bands respectively. The zigzag chain contains $N_y = 282$ AB sites for the Honeycomb lattice and $N_y = 426$ ABC sites for the dice lattice. (lower panel) Probability density distribution with the lattice sites along the finite direction corresponding to the edge modes shown in the upper panel, where the inset shows the localization and chirality of the modes along a rectangular slab. Parameter used are same as Fig.\ref{fig:band_structure_spin_polarized}}
    \label{fig:ebs_prob_density_delta}
\end{figure*}

\begin{figure*}
    \centering
    \includegraphics[scale=0.4]{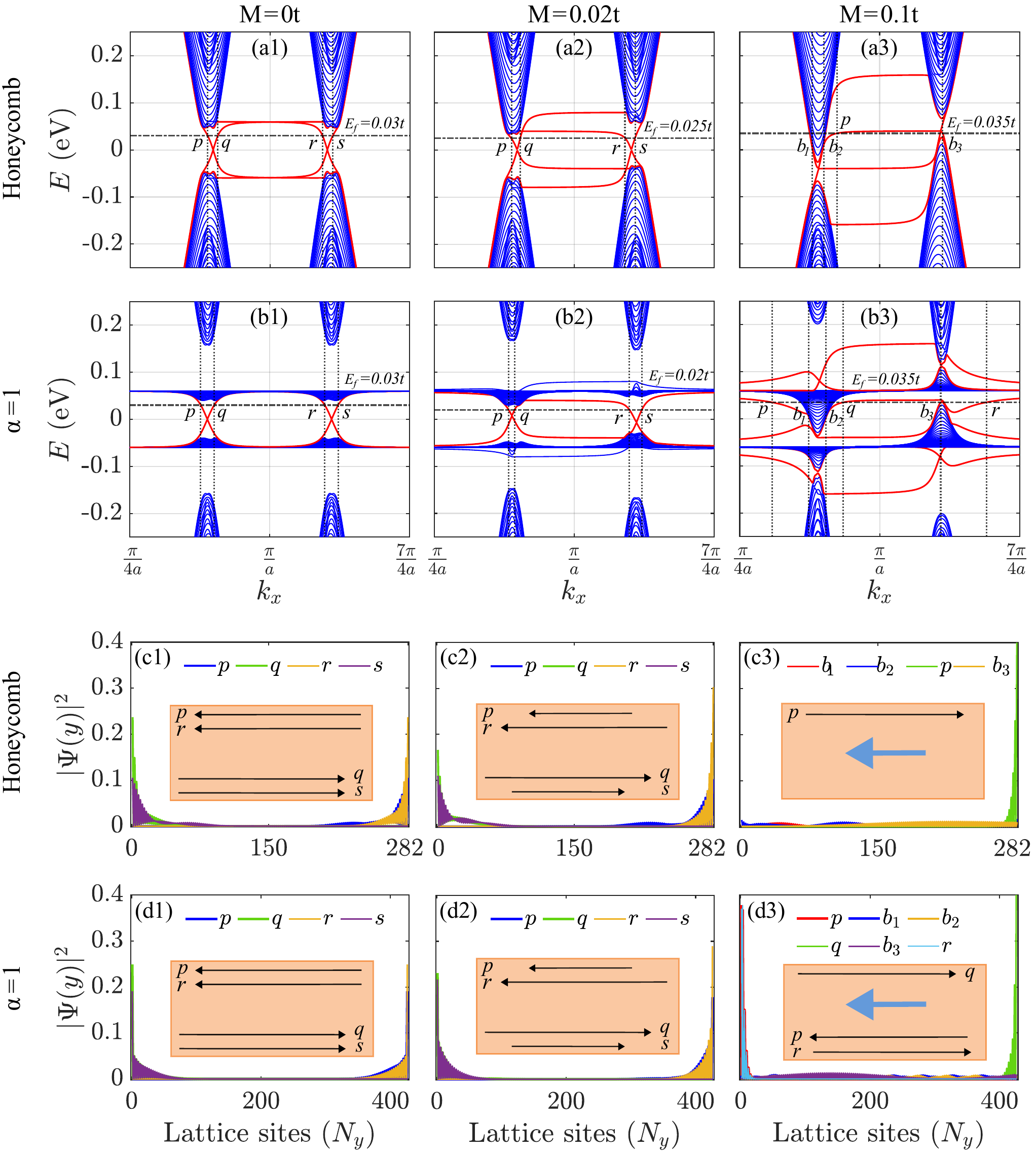}
    \caption{(upper panel) The band structures of zigzag nanoribbon with applied Rashba spin-orbit coupling and exchange term for Honeycomb lattice $[(a1) - (a3)]$ and dice lattice $[(b1) - (b3)]$ for three different values of staggered magnetization $\Delta M$, where red and blue color indicates the conducting edge bands and bulk bands respectively. The zigzag chain contains $N = 282$ AB sites for the Honeycomb lattice and $N = 426$ ABC sites for the dice lattice. (lower panel) Probability density distribution with the lattice sites along the finite direction corresponding to the edge modes shown in the upper panel, where the inset shows the localization and chirality of the modes along a rectangular slab.Parameter used are same as Fig.\ref{fig:band_structure_spin_polarized}}
    \label{fig:ebs_prob_density_M}
\end{figure*}
The concept of bulk-boundary correspondence states that topological phases possess localized edge states protected by nontrivial bulk topological invariants. The edge states in quantum anomalous Hall materials are generally associated with the total Chern number of the valence bands. Here we have discussed the edge states of the zigzag nanoribbon (ZNR) structure that explain the VPQAH phase and the TM phase individually obtained by two different staggered terms in the Hamiltonian, as discussed earlier in Sections \ref{sec: b} and \ref{sec: c}.

To plot the edge states, the band structure of the honeycomb and dice lattice nanoribbons are obtained by considering periodic boundary conditions along the direction of the zigzag edge and open boundary conditions along the perpendicular direction(see Fig.\ref{fig:lattice_schematic}).
For the honeycomb lattice, the ZNR contains sublattices A and B along opposite edges, whereas, for the Dice lattice, we have considered a C-A edged ribbon, i.e., where the top edge has A sublattices and the bottom edge contains C~\cite{Tan_2021,li2022novel}. The band structure for armchair nanoribbons are given in Appendix \ref{app:armchair_edge_bands}. Figs.\ref{fig:ebs_prob_density_delta}(a1) - \ref{fig:ebs_prob_density_delta}(a3) and Figs.\ref{fig:ebs_prob_density_delta}(b1) - \ref{fig:ebs_prob_density_delta}(b3) shows the ZNR band structure for different staggered electric potentials for honeycomb and Dice lattice, respectively with the Fermi energy crossings of the edge bands. The red and blue color denotes conducting edge bands and gapped bulk bands. The probability density distribution with the lattice sites along the perpendicular direction corresponding to edge modes for a given Fermi energy are shown in Figs.\ref{fig:ebs_prob_density_delta}(c1) - \ref{fig:ebs_prob_density_delta}(c3) and Figs.\ref{fig:ebs_prob_density_delta}(d1) - \ref{fig:ebs_prob_density_delta}(d3) where the insets suggest the localization and chirality of the edge modes in rectangular slab. Figs.\ref{fig:ebs_prob_density_delta}(a1) and \ref{fig:ebs_prob_density_delta}(b1) refer to edge bandstructure corresponding to bulk band shown in Figs.\ref{fig:band_structure_spin_polarized}(d) and \ref{fig:band_structure_spin_polarized}(h) where no external potential is applied. Two pairs of counter-propagating chiral edge modes are consistent with the total Chern number $C=2$ below the given Fermi energy $E_f = 0.03t$.
For a finite value of staggered electric potential ($\Delta=0.02t$) below the critical point, those two pairs of chiral edge states persist, however one pair of edge modes spread into the bulk and has a larger localization length due to a smaller bulk gap of one valley compared to the other valley [Figs.\ref{fig:ebs_prob_density_delta}(a2), \ref{fig:ebs_prob_density_delta}(b2) and Figs.\ref{fig:ebs_prob_density_delta}(c2), \ref{fig:ebs_prob_density_delta}(d2)]. The wave function of an edge state typically decays as $e^{-y/\gamma} \cos{kx}$, where $y$ is the distance from the edge, wave vector along the edge, and $\gamma$ is the localization length, which determines how fast the wave function decays into the bulk, and is inversely proportional to the band gap. 

After the critical point, there are no edge modes conducting from the valence band to the conduction band for the honeycomb lattice [Fig.\ref{fig:ebs_prob_density_delta}(a3)], hence turning into a bulk insulator. However, in the case of the dice lattice, now there is one pair of edge modes in each edge propagating in opposite directions  [Fig.\ref{fig:ebs_prob_density_delta}(b3)]. Notably, the contribution to the conducting edge modes comes from a specific valley, and their chirality is now altered. This valley-specific behavior gives rise to a valley-polarized QAH phase with a change in sign. In such a gapped system, the Fermi energy is chosen in the bulk gap that crosses only the conducting edge modes, which appear in a quasi-1D ribbon geometry. Therefore, the total Chern number of the bands below the Fermi energy includes all valence bands, and it corresponds to the total number of edge modes [Figs.\ref{fig:bulk_bs_bgap_chocc_delta}(c) and \ref{fig:bulk_bs_bgap_chocc_delta}(d)]. 

In the case of applied staggered magnetization, the system remains bulk-gapped as long as $M < \lambda_{ex}$ and the quantized number of edge modes propagate along the edges in opposite directions protected by the total Chern number $C=2$ as shown in Figs.\ref{fig:ebs_prob_density_M}(a2), \ref{fig:ebs_prob_density_M}(b2), \ref{fig:ebs_prob_density_M}(c2) and \ref{fig:ebs_prob_density_M}(d2). The bulk gap vanishes for $M > \lambda_{ex}$, and hence, the total Chern number of the valence bands does not correspond to the edge modes directly. 
For Fermi energy close to charge neutrality, there exist edge modes(red) [$p$ in Fig.\ref{fig:ebs_prob_density_M}(a3) and $p$, $q$ and $r$ in Fig.\ref{fig:ebs_prob_density_M}(b3)] propagating along the edges with a smaller localization length [Fig.\ref{fig:ebs_prob_density_M}(c3) and Fig.\ref{fig:ebs_prob_density_M}(d3)] along with conducting bulk modes (blue) [$b_1$, $b_2$ and $b_3$ in Fig.\ref{fig:ebs_prob_density_M}(a3) and \ref{fig:ebs_prob_density_M}(b3)]. The coexistence of edge modes and in-gap bulk modes are found recently in the modified Haldane model for both Honeycomb lattices~\cite{PhysRevLett.120.086603, PhysRevB.99.115423, PhysRevB.107.045117, PhysRevB.101.214102} and $\alpha$-$\mathcal{T}_3$ lattice~\cite{PhysRevB.109.235105, PhysRevB.111.045406} where co-propagating edge modes exist in the two opposite edges along with modes propagating in the bulk in the opposite direction. This pair of co-propagating edge modes is known as anti-chiral edge states and are found in a topological metal phase of matter, which has also been recently experimentally realized~\cite{PhysRevLett.125.263603, cheng2023revealing}. In our case, the Honeycomb lattice hosts one unpaired edge mode $p$ propagating only on one edge of the ribbon [Fig.\ref{fig:ebs_prob_density_M}(c3)] and compensated by the counter-propagating bulk. Whereas, for the dice lattice edge modes $p$ and $r$ being on the same edge and counter-propagating, cancels each other, resulting into a single edge mode $q$ propagating on the other edge, which makes it equivalent to the honeycomb case. These unpaired edge states, referred to as half-antichiral edge states (HACES), were recently proposed in a composite Haldane-modified Haldane bilayer system of Honeycomb lattice~\cite{PhysRevB.110.165303}.

\subsection{Anomalous Hall conductance}
In this section, we calculate the anomalous Hall conductivity for the Honeycomb and Dice lattice with RSOC and EC, with two individual staggered terms as described in Secs.\ref{sec: b} and \ref{sec: c}.
We computed the Hall conductivity by numerically integrating the Berry curvature (BC) of all occupied electronic states over the entire Brillouin zone (BZ). This involves summing up all the filled bands in the system~\cite{RevModPhys.82.1959}.
\begin{equation}
    \sigma_{xy} = \frac{\sigma_0}{2\pi} \sum_n \int \Omega_n (k_x,k_y) f(E_{k_x,k_y}^n) dk_x dk_y
\end{equation}
Where $\Omega_n$ is the berry curvature for $n^{th}$ band from Eq.(\ref{eq:bc_equation}), $f(E) = 1/[1+e^{(E-E_f)/{K_BT}}]$ is the Fermi-Dirac distribution function where $E_f$ and $T$ signifies Fermi energy and absolute temperature respectively, $E_{k_x,k_y}^n$ is the energy eigenvalues for $n^{th}$ band and $\sigma_0 = e^2/h$. 
\begin{figure}[!ht]
    \centering
    \includegraphics[scale=0.33]{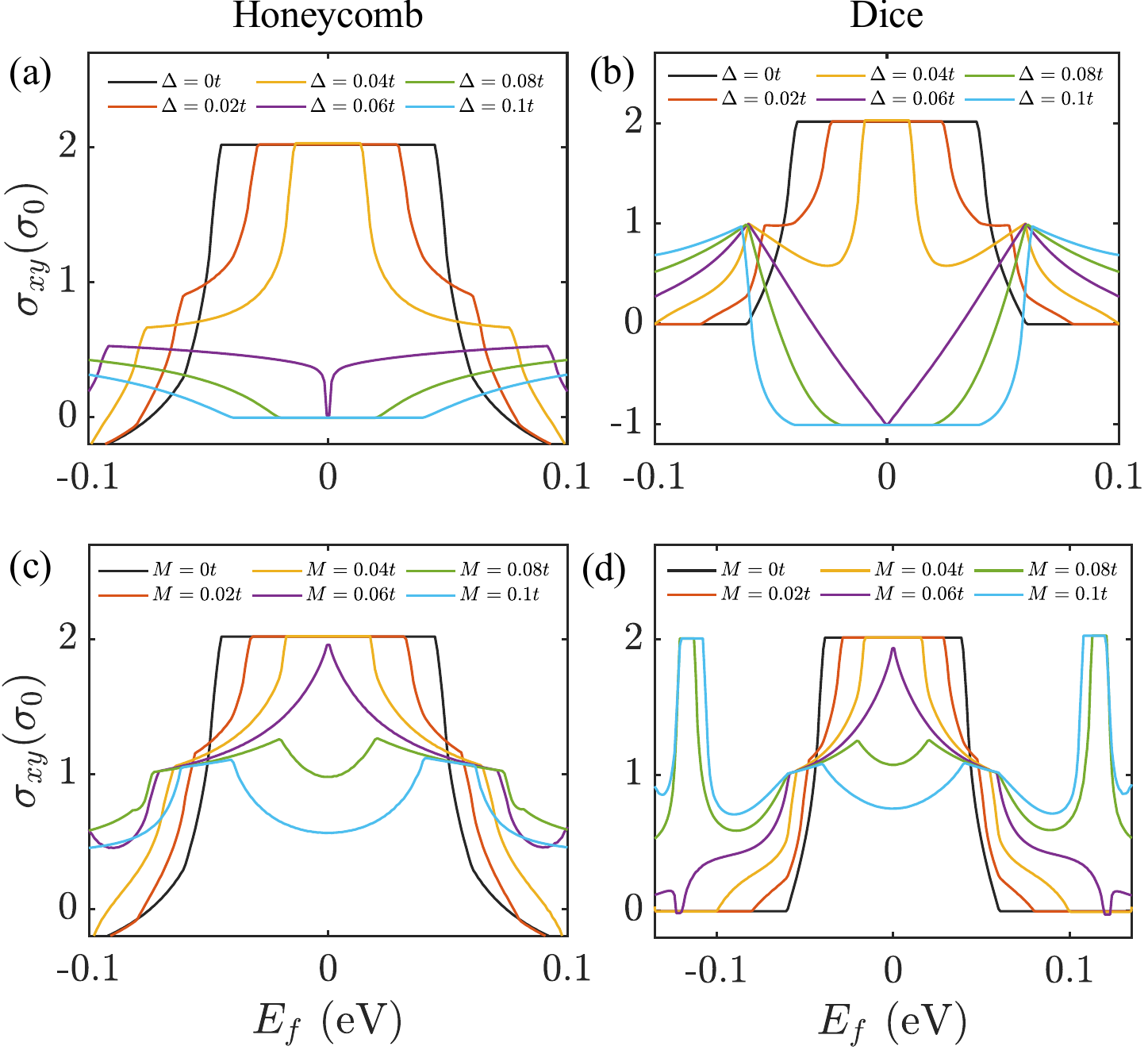}
    \caption{QAHC for different values of [upper panel] staggered electric potential $\Delta$ for Honeycomb lattice (a) and Dice lattice (b) and [lower panel] staggered magnetization $M$ for Honeycomb lattice (c) and Dice lattice (d) as a function of Fermi energy.}
    \label{fig:qahc_collage}
\end{figure}

When the Fermi energy lies in a band gap, the Fermi-Dirac distribution function $f(E)$ at zero absolute temperature equals unity, and the total contribution of all occupied states comes from bands below the Fermi energy. Then, the integration over the BZ gives the total Chern number of the bands below the Fermi energy, and we get a plateau $\sigma_{xy} = |C_n|e^2/h$. The conductivity decays when the Fermi energy lies outside the bulk gap or when the gap closes. 

Fig.\ref{fig:qahc_collage}(a) and \ref{fig:qahc_collage}(c) shows the anomalous Hall conductivity for the Honeycomb lattice and Dice lattice as a function of Fermi energy ($E_f$) in the unit of $\sigma_0$ for different values of staggered electric potential $\Delta$. 
The width of the Hall plateau decreases with the gap ($\Delta E_{direct}^{h(d)}$), and the plateau completely diminishes when the gap is zero at $\Delta=\lambda_{ex}$. As the gap reopens at $K^\prime$ a new quantized plateau arises, suggesting a QAH phase transition. The Hall plateau changes from $2e^2/h$ (QAH phase) to $0$ (BI phase) for the honeycomb lattice and from $2e^2/h$ (QAH phase) to $-e^2/h$ (VPQAH phase), where the negative sign refers to a change in the chirality of the edge modes. 
Fig.\ref{fig:qahc_collage}(c) and \ref{fig:qahc_collage}(d) show the QAHC as a function of Fermi energy for different values of staggered magnetization $M$. 
In this case, the indirect bulk band gap $\Delta E_{indirect}^{h(d)}$ decreases with increasing $M$ and vanishes at $M = \lambda_{ex}$ and so the QAH plateau.
For $M>\lambda_{ex}$, the system remains metallic, and the Hall conductivity is no longer quantized. In the TM phase, the non-zero contribution to the QAHC for Fermi energy close to the charge neutrality arises from the half anti-chiral edge states described previously. 
Fig.\ref{fig:qahc_collage}(d) shows two more smaller plateau regions for the Fermi energy away from charge neutrality. Those two plateaus change from $0$ to $2e^2/h$ at the band crossing between VB1 and VB2 at valley-$K$ and between CB1 and CB2 at valley-$K^\prime$.
When Fermi energy lies in the region of a small gap between VB2 and VB3, the AHC is solely contributed by VB1 and VB2, resulting in $\sigma_{xy}=0$ and $\sigma_{xy}=|2|e^2/h$ respectively before and after the band crossing (Fig.\ref{fig:qahc_collage}(d)), as the total Chern number $C=2$ is the sum of the two Chern numbers of VB2 and VB3. Similarly, we observe another plateau region for the energy gap between CB1 and CB2 as the Chern numbers of CB1 and VB1 are equal and opposite signs, they cancel out, resulting in the total Chern number being the same as that in the valence band gap.
\begin{figure*}[!ht]
    \centering
    \includegraphics[scale=0.3]{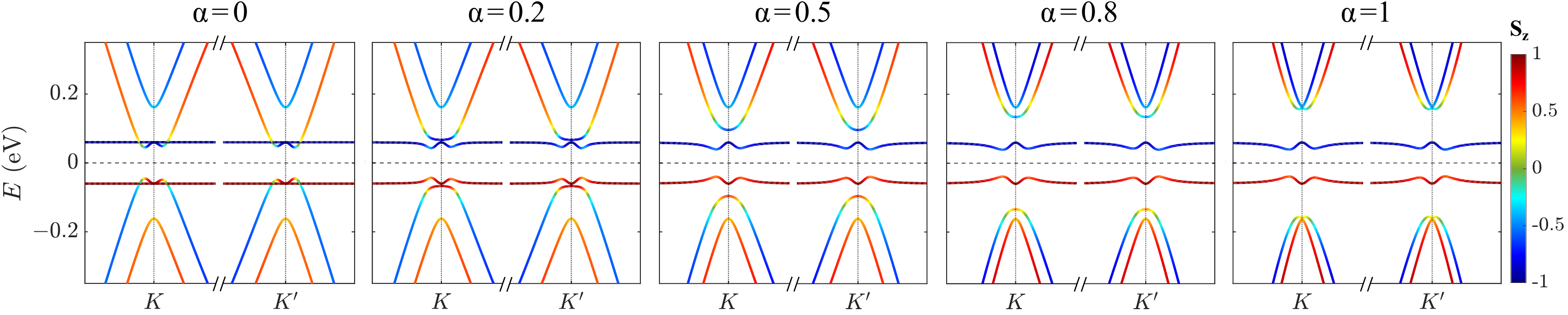}
    \caption{Band structure obtained from $H$ in Eq.(\ref{eq:Full_hamiltonian}) near the two inequivalent $K$-points for different values of $\alpha$, for hamiltonian parameters used in terms of $t$ are, $\lambda_R = 0.05$, $\lambda_{ex} = 0.06$. The color of the bands represents the expectation value of the z-component of the spin.}
    \label{fig:bands_intermediate_alpha}
\end{figure*}
\begin{figure}[!ht]
    \centering
    \includegraphics[scale=0.43]{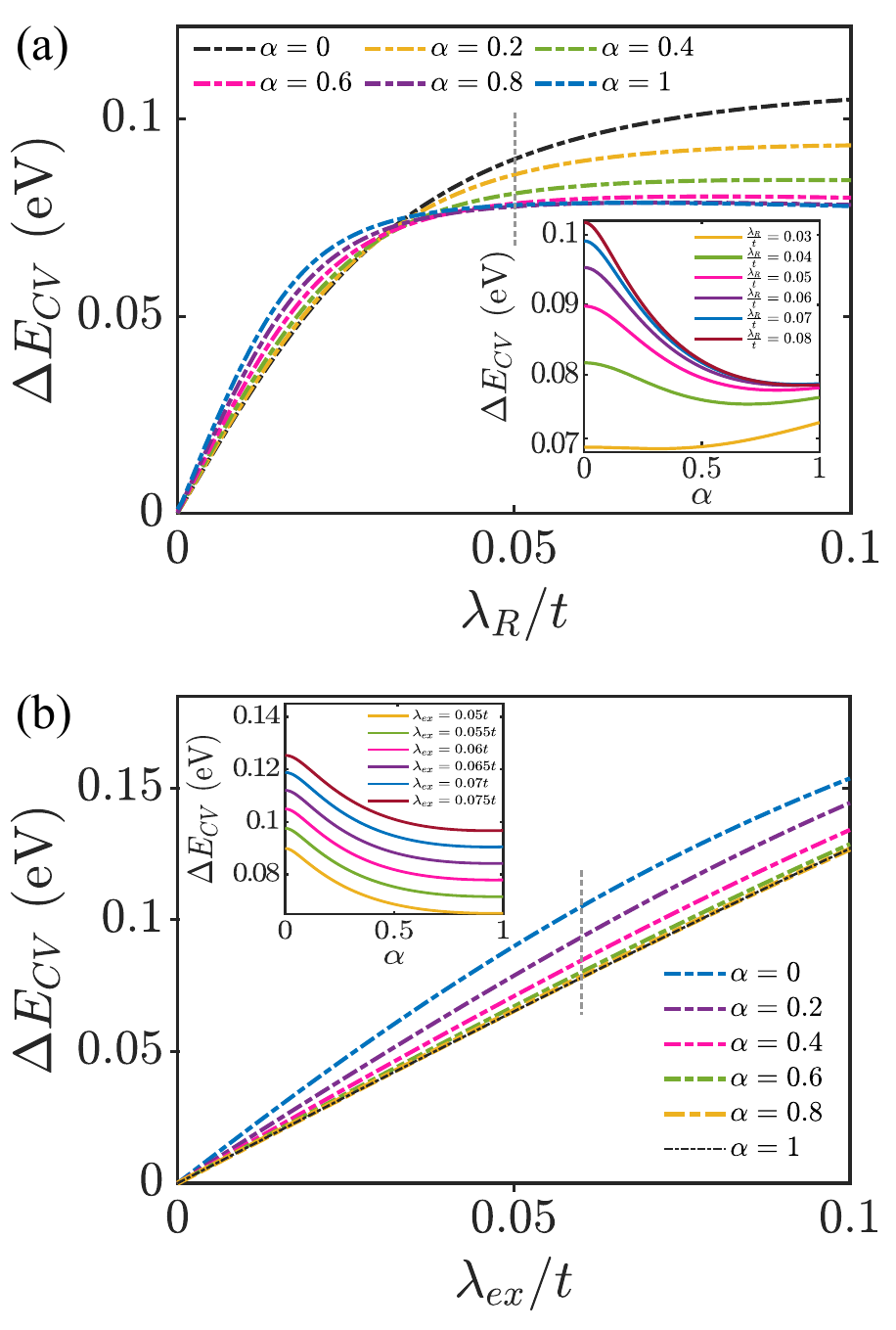}
    \caption{Band gap vs (a) Rshba SOC strength and (b) exchange coupling for different values of $\alpha$. The fixed parameters are kept the same as in Fig.\ref{fig:band_structure_spin_polarized}.}
    \label{fig:bgap_rashba_ex_alpha}
\end{figure}
\begin{figure*}[!ht]
    \centering
    \includegraphics[scale=0.38]{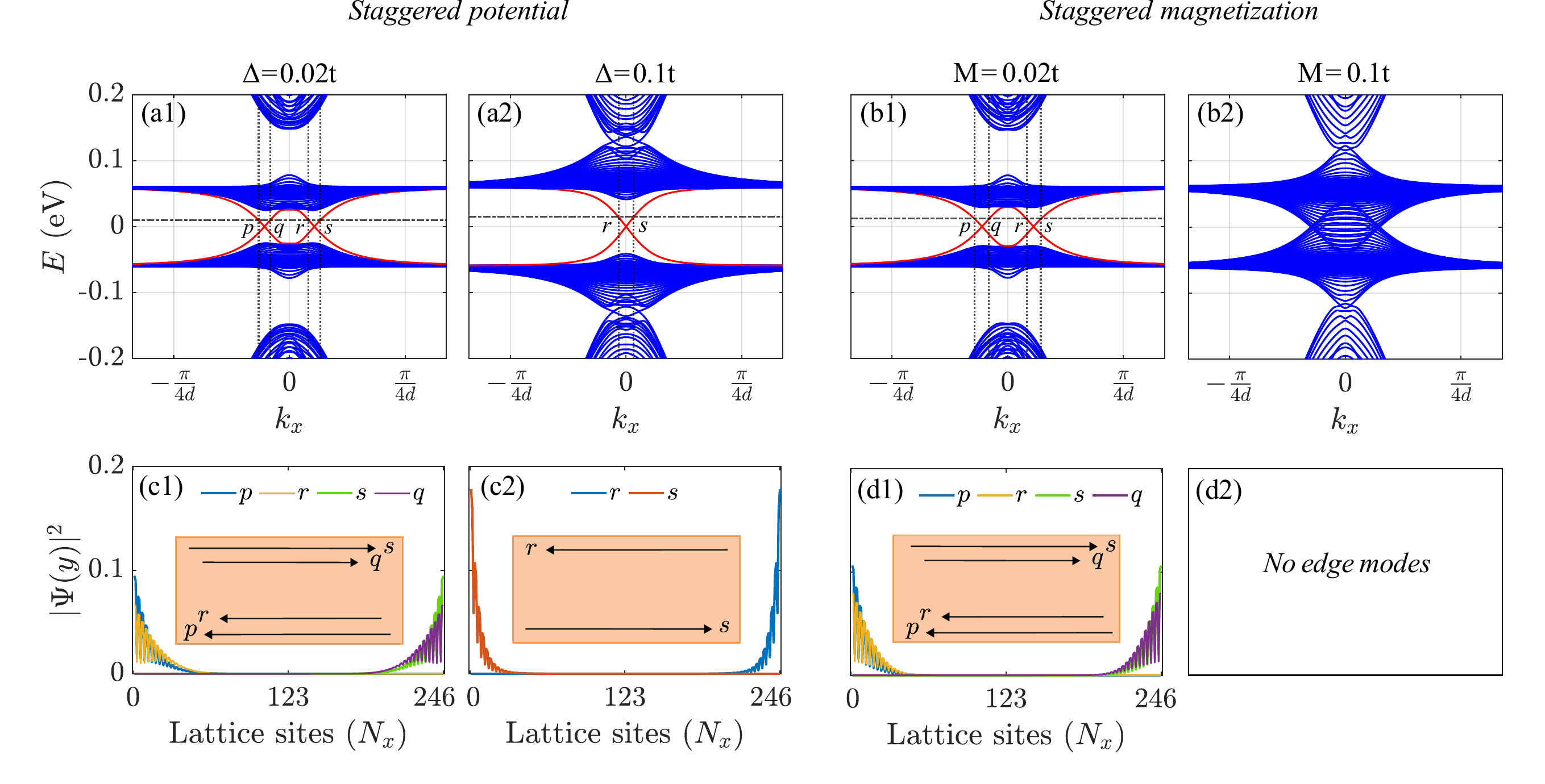}
    \caption{(upper panel) The band structure of Dice lattice armchair nanoribbon for staggered potential $\Delta < \lambda_{ex}$ (a1) and $\Delta > \lambda_{ex}$ (a2) and, staggered magnetization $M < \lambda_{ex}$ (b1) and $M > \lambda_{ex}$ (b2). (lower panel) The probability densities along the width corresponding to edge modes in the upper panel. The width of the nanoribbon is taken as $123\sqrt{3}a_0$ along the $x$-direction. Other parameters used are same as that of Fig.\ref{fig:band_structure_spin_polarized}}
    \label{fig:armchair_ebs_prob_density}
\end{figure*}
\section{Conclusion}\label{sec:conclusions}
We have investigated the electronic band structures and topological properties of a gapped phase obtained by the interaction of Rashba SOC and exchange coupling for a pseudospin-$1$ system, $\alpha$-$\mathcal{T}_3$ lattice with $\alpha=1$, and compared the results with that of a pseudospin-$1/2$ system, Honeycomb lattice. We also uncover the effect of external staggered electric potential and magnetization on these topological phases individually. We observed a topological phase transition from the QAH phase ($2e^2/h$) to a bulk insulator for Honeycomb ($\alpha=0$) and to a valley-polarized QAH phase ($-e^2/h$) with a flip in chirality/edge for Dice lattice $(\alpha=1)$ at a staggered electric potential equal to the exchange coupling.
For an applied staggered magnetization, the Chern number is insufficient to characterize the topological phase of the system as it experiences a phase transition from QAH ($2e^2/h$) phase to a TM phase at $ M = \lambda_{ex}$. The TM phases in both Honeycomb and Dice lattice host a single edge mode along one edge, referred to as a half-antichiral edge state, along with counter-propagating bulk modes.
In the TM phase, the bulk gap closes, the system becomes metallic, and QAH is no longer quantized. However, contrary to the honeycomb lattice, for Fermi energy away from the bulk gap, a pair of new quantized plateaus ($2e^2/h$) arise for the dice lattice. 

\begin{acknowledgments}
We acknowledge the support provided by the Kepler Computing facility, maintained by the Department of Physical Sciences, IISER Kolkata, for various computational needs. P.P. acknowledges support from the Council of Scientific and Industrial Research (CSIR), India, for the doctoral fellowship. B.L.C. acknowledges the SERB with grant no. SRG/2022/001102 and ``IISER Kolkata Start-up-Grant" Ref. No. IISER-K/DoRD/SUG/BC/2021-22/376.
\end{acknowledgments}

\appendix
\section{Beyond Dice lattice : \texorpdfstring{$0 < \alpha <1$}{0 < alpha <1}} \label{app:intermediate_alpha}
In this manuscript, so far we have discussed the QAH phase induced by RSOC and exchange coupling on dice lattice, i.e., $\alpha=1$ case of the model discussed in sec.\ref{sec:model_discussion}. The parameter hopping $\alpha$ in $\alpha$-$\mathcal{T}_3$ lattice is known to tune the topology of the system with Haldane and modified Haldane model~\cite{PhysRevB.109.235105}, intrinsic SOC~\cite{PhysRevB.103.075419}, or Floquet engineering~\cite{PhysRevB.111.045406}. However, we observed that the topological phase obtained by RSOC and exchange term remain unchanged irrespective of $\alpha$ values. Fig.\ref{fig:bands_intermediate_alpha} shows the band structures of the gapped phase ($\lambda_R \neq 0$, $\lambda_{ex} \neq 0$) obtained from Eq.(\ref{eq:Full_hamiltonian}) for intermediate values of alpha including two limiting values $\alpha=0$ (honeycomb lattice) and $\alpha=1$ (dice lattice), where the dashed line indicates the Fermi energy and the color of the bands represents the expectation values of the z-component of spin.

At $\alpha=0$, the $6\times6$ spinful Hamiltonian mimics the $4\times4$ Hamiltonian of a Honeycomb lattice, apart from the two completely flat bands separated by $2\lambda_{ex}$ arising from the onsite exchange coupling term of the isolated $C$ sublattice. These two bands, being completely nondispersive, do not contribute to the Berry curvature and conductivity of the system. For a finite nonzero value of $\alpha$, the middle bands become dispersive; however, the bulk gap remains open for any $\alpha$. Therefore, although the band gap varies, no new topological phase arises. 
The variation in the band gap $\Delta E_{CV}$ with Rashba SOC strength and exchange coupling for different $\alpha$ are shown in Figs.\ref{fig:bgap_rashba_ex_alpha} (a) and \ref{fig:bgap_rashba_ex_alpha} (b), where the insets show the band gap plot with alpha for different RSOC and exchange coupling, respectively. For any specific $\alpha$, the band gap opens when both $\lambda_R$ or $\lambda_{ex}$ are non-zero. For lower (higher) values of Rashba coupling, the band gap increases (decreases) with $\alpha$. However, it always increases with the exchange coupling.  The gray lines in the band gap plots indicate the parameters and the gap corresponding to different $\alpha$ for the band structure shown in Fig.\ref{fig:bands_intermediate_alpha}. 
\section{Armchair edge bands}\label{app:armchair_edge_bands}
To obtain the band structure of the armchair nanoribbon, the periodic boundary condition along the $y$-axis (armchair edge) and open boundary condition along the $x$-direction (zigzag edge) are considered (see Fig.\ref{fig:lattice_schematic}). The upper panel of Fig.\ref{fig:armchair_ebs_prob_density} shows the band structure of armchair nanoribbons of dice lattice ($\alpha=1$) and the Fermi energy crossings of the edge bands, and the lower panel shows the probability density distribution across the lattice sites along the direction of finite width, with inset showing schematic of chirality of edge modes in a rectangular slab. For the staggered potential, the Fermi energy crosses two (one) pair of edge bands before (after) $\Delta = \lambda_{ex}$, consistent with the HCI (CI) phase [Figs.\ref{fig:armchair_ebs_prob_density}(a1) and \ref{fig:armchair_ebs_prob_density}(a2)]. It is also important to note that the reversal of the chirality of edge modes after $\Delta = \lambda_{ex}$ is also consistent with that of the zigzag edge and flip in the sign of QAH. However, since the armchair geometry overlaps the two valleys, the valley polarization is not evident in contrary to the zigzag edge. For the staggered magnetization, before $M = \lambda_{ex}$, the Fermi energy crosses two pairs of edge bands and hence is a HCI phase; however, for $M > \lambda_{ex}$ the conduction bulk states overlap with the valance band bulk states and no edge bands are observed thus appearing as a trivial metal [Figs.\ref{fig:armchair_ebs_prob_density}(b1) and \ref{fig:armchair_ebs_prob_density}(b2)], in contrary to the zigzag edge bands. This is because the edge bands for TM phase in Fig.\ref{fig:ebs_prob_density_M} (b3) connecting the valence band of $K$-valley to the conduction band of $K^\prime$-valley and shielded from the view by the in-gap bulk bands as soon as the indirect band gap is closed. Hence, valley separation is crucial to visualize these half anti-chiral edge states.

\nocite{*}

\bibliography{apssamp}

\end{document}